\documentclass[12pt]{article}
\usepackage{amsmath, graphicx, bm,amsfonts, amsthm,ctable,savefnmark, subfigure}
\usepackage{color,setspace}
\usepackage[numbers]{natbib}
\usepackage{float}
\newcommand{\btable}[3]{
\begin{table}[htbp]
\begin{center}
\caption{#2\label{#3}}
\begin{tabular}{#1}}
\newcommand{\etable}{\end{tabular}
\end{center}
\end{table}}

\newcommand{\mc}[2]{\multicolumn{#1}{c}{#2}}

\newtheorem{theorem}{Theorem}
\newtheorem{corollary}{Corollary}
\theoremstyle{definition}

\DeclareMathOperator{\E}{E} \DeclareMathOperator{\var}{var}
\DeclareMathOperator{\bias}{bias} \DeclareMathOperator{\MSE}{MSE}
\newenvironment{problist}
   {
      \begin{list}
         {---}
         {
            \setlength{\itemsep}{.5ex}
            \setlength{\parsep}{0ex}
            \setlength{\parskip}{0ex}
            \setlength{\topsep}{.5ex}
         }
   }
   {
      \end{list}
   }

\doublespace

\begin{document}

\date{}

\title{CDF and Survival Function Estimation with \\Infinite-Order Kernels\footnote{Running head: Reduced-Bias CDF Estimation}}

\author{Arthur Berg\footnote{Corresponding author} \footnote{Department of Statistics; University of Florida; 408 McCarty C; Gainesville, FL  32611-0339, USA;
{\texttt berg@ufl.edu}}
\ and Dimitris N. Politis\footnote{Department of Mathematics; University of
California, San Diego; La Jolla, CA 92093-0112,  USA; {\texttt
dpolitis@ucsd.edu}}}

\maketitle

\abstract{A reduced-bias nonparametric estimator of the cumulative distribution function (CDF) and the survival function is proposed using infinite-order kernels.  Fourier transform theory on generalized functions is utilized to obtain the improved bias estimates. The new estimators are analyzed in terms of their relative deficiency to the empirical distribution function and Kaplan-Meier estimator, and even improvements in terms of asymptotic relative efficiency (ARE) are present under specified assumptions on the data.  The deficiency analysis introduces a deficiency \emph{rate} which provides a continuum between the classical deficiency analysis and an efficiency analysis.  Additionally, an automatic bandwidth selection algorithm, specially tailored to the infinite-order kernels, is incorporated into the estimators.  In small sample sizes these estimators can significantly improve the estimation of the CDF and survival function as is illustrated through the deficiency analysis and computer simulations.
 }\vspace{.5cm}



\section{Introduction}

We consider the problem of estimating the CDF in contexts of independently and identically distributed (iid) data and randomly right-censored data. Indeed, the seminal paper of Kaplan and Meier \cite{kaplan58} solves this problem with the product-limit estimator---the nonparametric maximum likelihood estimator of the CDF---but there is still room for improvement, especially when the sample size is small.

The most obvious drawback of the Kaplan-Meier estimator, like the empirical distribution function (EDF), is its lack of smoothness.  Kernel smoothing easily remedies this problem, but also introduces two new issues of choosing the best kernel and bandwidth.  Kernel smoothing also improves the estimator mean square error (MSE) performance by decreasing its variance while introducing a slight bias resulting in an overall improvement of the MSE.  The MSE improvement, however, is typically only a second-order improvement, since the original estimator's first-order MSE convergence rate already achieves the best-possible $\sqrt{n}$-rate.  When the asymptotic relative efficiency (ARE) between the Kaplan-Meier estimator and its smoothed counterpart is one, as is typically the case, a distinction in performance can be measured by considering the asymptotic relative \textit{deficiency}, or just simply the deficiency between the two estimators.  The general notion of deficiency and subsequent calculations with the proposed estimators is provided in Section \ref{section:deficiency} which also illustrates that an actual increase in efficiency can be achieved with the new estimators under certain (rather strong) assumptions of the distribution function.

Higher-order MSE improvement is influenced by the kernel order---the higher the kernel order, the greater the improvement.  Therefore the best kernel-based estimators, the ones with smallest asymptotic MSE, are
the estimators that use infinite-order kernels.  Current methods traditionally invoke second-order kernels \cite{reiss81} and more recently a hybrid kernel estimator has been investigated \cite{kim06}, but infinite-order kernel methods allow for the greatest improvement in bias rates without affecting the rates of the variance.  The main argument against the use of large-order kernels in \emph{density} estimation is the concern that the estimator may be negative on some intervals when it is known that the true probability density is always nonnegative.  This argument, however, is moot in the density estimation context (so also in the CDF estimation context) since the estimator can easily be truncated to zero when it goes negative then renormalized to have a total area of one without affecting the MSE convergence rate.  General construction of the infinite-order kernel estimators are introduced in the following section and a compatible bandwidth selection algorithm that adapts to the infinite-order kernels is described in Section \ref{section:bandwidth}.

Another pitfall of all kernel estimators of the density is the lack of consistency at boundary points when the support of the density lies in an interval or half-interval.  Simple reflection \cite{silverman1986des} solves this problem in the density estimation context and an analogous fix also exists for CDF estimators.  Boundary correction and standardization methods specific to kernel-smoothed CDF estimators are discussed in Section \ref{section:boundary}.

Simulations with iid and censored data illustrate the effectiveness of the infinite-order kernel estimators coupled with the automatic bandwidth selection algorithm of Section \ref{section:simulations}.   Uniform improvement in MSE over existing estimators is observed in the simulations.  Since estimation of the CDF is so fundamental in standard statical analysis, there are many applications of the new estimators beyond just estimating the underlying CDF.  Some of these applications are included in the last section on Discussions and Conclusions.

\section{Estimation with Flat-Top Kernels}
\label{section:estimators}
The analysis will be confined to independently and identically distributed (iid) data, but extensions to randomly right censored with possible left truncation can be more generally derived; cf. \cite{Berg07,cao99}.

Let $X_1,\ldots,X_n$ be independent\footnote{The independent assumption can be relaxed under certain stationarity and mixing conditions; see \cite{Liu06,gyorfi89}.} and identically distributed random vectors in $\mathbb{R}$ with absolutely continuous distribution function $F$ and corresponding probability density function $f$. Estimation of $f$ with infinite-order kernels was considered in \cite{Politis99} and \cite{Berg07}; here we consider the integration of those estimators in the construction of the CDF estimator.

The traditional estimator of the CDF is the empirical distribution function, or EDF, which is given by
\[
\hat{F}(t)=\frac{1}{n}\sum_{j=1}^n I(X_i\le t)
\]
where $I(\cdot)$ represents the indicator function.  The kernel estimator of the probability density, $f$, is then given by
\[
  \hat{f}_{h}(x)=\int_{-\infty}^\infty \frac{1}{h}K \left(\frac{t-X_j}{h}\right)\,d\hat{F}(t)=
  \frac{1}{n h}\sum_{j=1}^{n}
  K\left(\frac{x-X_j}{h}\right).
\]
where $K$ is a kernel that integrates to one (but not necessarily nonnegative!) and $h$ is the bandwidth parameter.  To insure consistency of $\hat{f}_h$, $h$ should satisfy the condition $h\rightarrow0$ as $n\rightarrow\infty$ but with $nh\rightarrow \infty$.

The smoothed estimation of the CDF, $\hat{F}_{h}$, is
constructed by integrating $\hat{f}_{h}$.  That is,
\begin{equation}
\label{eq:Fhat} \hat{F}_{h}(t)=\int_{-\infty}^{t}
\hat{f}_{h}(x)\,dx=\frac{1}{n}\sum_{j=1}^n
\bar{K}\left(\frac{t-X_j}{h}\right)
\end{equation}
where $\bar{K}(t)=\int_{-\infty}^{t}K(x)\,dx$.

The estimator $\hat{F}_{h}(t)$ is equivalent to the EDF in terms of first-order asymptotic performance, but improvements are achieved in the higher-order terms.  The estimator $\hat{F}_{h}(t)$ effectively
smooths the EDF, decreasing its variance at the cost of introducing a slight bias.  The variance improvement is uniform across different kernels, affecting only the second-order constant and not the
second-order \emph{rate} (refer to equation \eqref{eq:var} below); however the additional bias that gets introduced in the smoothing can be minimized significantly by using kernels of large order with
infinite-order kernels providing the most benefit.  The variance of $\hat{F}_h(t)$, as derived in \cite{li07}, is given by
\begin{equation}
\label{eq:var} \var\left[\hat{F}_h(t)\right]=\frac{F(t)[1-F(t)]}{n}
-2 f(t) \left(\int
u\bar{K}(u)K(u)\,du\right)\frac{h}{n}+o\left(\frac{h}{n}\right).
\end{equation}
The bandwidth parameter $h$ only enters the variance expression through the second-order term which is negative.  So the larger $h$ is, the smaller the variance of $\hat{F}_h(t)$ becomes.  However, we will
see below in Theorem \ref{thm:bias} that the smaller $h$ is, the smaller the bias of $\hat{F}_h(t)$ becomes.  Therefore there is an optimal $h$ that strikes a compromise between the bias and variance terms which is presented in Corollary \ref{cor:bias} below.

We now construct a family of infinite-order kernels, following \cite{Politis99}, that are derived from ``flat-top functions''. We start with a continuous, real-valued function $\kappa$ given by
\begin{equation}
\label{eq:kappa}
\kappa(s)=
\begin{cases}
  1, & |s|\le c\\
  g(|s|), & \text{otherwise}
\end{cases}
\end{equation}
where $g$ is any continuous, square-integrable function that is bounded in absolute value by one and satisfies $g(|c|)=1$.  The region $|s|<c$ is referred to as the ``flat-top neighborhood'', but in some cases we may wish to relax the requirement to allow $g(s)\approx 1$ when $s$ is close to $c$.  This ``effective flat-top neighborhood'' is useful when using an infinitely smooth function $\kappa(s)$ as described in \cite{politis07} and Section 6 below.  The Fourier transform of $\kappa$ then produces the infinite-order kernel, $K$, of interest. Specifically,
\begin{equation}
\label{eq:infinite}
K(x)=\frac{1}{2\pi}\int_{-\infty}^\infty \kappa(s) e^{-i s x} \, ds.
\end{equation}

The MSE of $\hat{F}_{h}(t)$ with an infinite-order kernel $K$ is now computed under various assumptions on the smoothness of the underlying density.  Let $\phi(t)$ be the characteristic function
corresponding to $f(x)$, i.e.
\[
\phi(s)=\int_{-\infty}^\infty f(x) e^{i s x}\,dx.
\]
The following three assumptions quantifies the degree of smoothness of the density $f(x)$ by the rate of decay of its characteristic function.
\begin{description}
  \item[Assumption $A(p)$:]  There is a $p>0$ such that   $\int_{-\infty}^\infty |t|^p\,|\phi(t)|<\infty$.
  \item[Assumption $B$:]  There are positive constants $d$ and $D$ such that $|\phi(t)|\le D e^{-d|t|}$.
  \item[Assumption $C$:]  There is a positive constant $b$ such that $\phi(t)=0$ when $|t|\ge b$.
\end{description}

\begin{theorem}
\label{thm:bias}
Let $\hat{F}_h(t)$ be a kernel smoothed estimator of the CDF with an infinite-order kernel derived from a flat-top function.
\begin{problist}
\item[(i)]  Suppose assumption $A(p)$ holds, then
\[
\sup_{t\in\mathbb{R}}\left|\bias\left(\hat{F}_h(t)\right)\right|=o\left(h^{p+1}\right).
\]
\item[(ii)]  Suppose assumption $B$ holds, then
\[
\sup_{t\in\mathbb{R}}\left|\bias\left(\hat{F}_h(t)\right)\right|=O\left(he^{-d/h}\right)=o\left(e^{-d/h}\right).
\]
\item[(iii)]  Suppose assumption $C$ holds.  When $h\le 1/b$,
\[
\sup_{t\in\mathbb{R}}\left|\bias\left(\hat{F}_h(t)\right)\right|=0.
\]
\end{problist}
\end{theorem}

To optimize the amount of smoothing under the MSE criterion---i.e., to optimize the bandwidth $h$---we choose the bandwidth that allows the squared bias rates to be comparable to the second-order variance rates. The optimal bandwidths are provided in the following corollary.

\begin{corollary}
\label{cor:bias} Let $\hat{F}_h(t)$ be as in Theorem \ref{thm:bias}.
\begin{problist}
  \item[(i)]  Suppose assumption $A(p)$ holds.  Letting $h\sim an^{-\beta}$ where $a$ is any positive constant and $\beta=(2p+1)^{-1}$ optimizes the tradeoff between the bias and variance of $\hat{F}_h(t)$ and gives
\[
\sup_{t\in\mathbb{R}}\left|\bias\left(\hat{F}_h(t)\right)\right|=o\left(n^{-\frac{p+1}{2p+1}}\right).
\]
  \item[(ii)]  Suppose assumption $B$ holds.  Letting $h\sim a/\log n$ where $a<2d$ is a constant optimizes the tradeoff between the bias and variance of $\hat{F}_h(t)$ and gives
\[
\sup_{t\in\mathbb{R}}\left|\bias\left(\hat{F}_h(t)\right)\right|=o\left(\frac{1}{\sqrt{n}\log n}\right).
\]
\item[(iii)]
Suppose assumption $C$ holds.  Letting $h\le 1/b$ be fixed guarantees zero bias and the best possible variance rate.
\end{problist}
\end{corollary}

Estimation of the survival function with randomly right censored data can be similarly improved with the smoothing of the Kaplan-Meier estimator with infinite-order kernels.  Density estimation of censored data with infinite-order kernels is analyzed in \cite{Berg07}, and an estimator of the survival function can be similarly derived from this density estimator through integration as in \eqref{eq:Fhat}.  The same conclusions as Theorem \ref{thm:bias} and Corollary \ref{cor:bias} will also hold for the smoothed version of the Kaplan-Meier estimator with infinite-order kernels.  This is detailed in the following theorem where the proof has been omitted as it follows naturally from the iid case above.

Define $\hat{S}_h(t)$ to be a smoothed estimator of the survival function, $S(t)=1-F(t)$, derived from smoothing the Kaplan-Meier estimator with an infinite-order kernel of the form given in \eqref{eq:infinite}; i.e.,
\begin{equation}
\label{eq:smoothedKM}
\hat{S}_h(t)=\sum s_j \bar{K}\left(\frac{t-X_j}{h}\right)
\end{equation}
where $s_j$ is the height of the jump of the Kaplan-Meier estimator at $X_j$ (cf. \cite{Berg07} for more details).  The following theorem is consistent with the results described in \cite{kulasekera01}.
\begin{theorem}
Let $\hat{S}_h(t)$ be a kernel smoothed estimator of the survival function as in \eqref{eq:smoothedKM} above. Suppose assumption $A(p)$ holds, then
\[
\sup_{t\in\mathbb{R}}\left|\bias\left(\hat{S}_h(t)\right)\right|=o\left(h^{p+1}\right)=o\left(n^{-\frac{p+1}{2p+1}}\right)
\]
when $h\sim an^{-\beta}$ where $a$ is any positive constant and $\beta=(2p+1)^{-1}$.

\end{theorem}

The analysis under assumptions B and C of the above theorem are considerably more complex and have been omitted.

\section{Deficiency}
\label{section:deficiency}
The notion of deficiency was introduced in the article ``Deficiency'' by Hodges and Lehmann \cite{hodges70} wherein several deficiency calculations are provided.  Many articles followed suit using the deficiency concept to compare kernel-smoothed estimators, but many of the approaches used in calculating the deficiency strayed from the original and simple techniques employed by Hodges and Lehmann; c.f. \cite{azzalini81,falk83,ghorai89,ghorai90,reiss81,xiang95}.  The simplicity of the original deficiency computations is maintained in the proof of Theorem \ref{thm:power} below.

The deficiency concept is described as follows.  Given an estimator, $S_m$, based on a sample of size $m$ and a more efficient estimator, $T_n$, based on a sample of size $n$ with equivalent performance as $S_m$.  The difference between the sample sizes, $d=m-n$, defines the relative deficiency between the two estimators.  The original paper of Hodges and Lehmann mostly dealt with situations where $d$ approaches a finite limit as $n$ goes to infinity in which case the two estimators have an asymptotic relative \emph{efficiency} (ARE) of one.   However, it is still possible for two estimators to have an ARE of one yet with a deficiency that approaches infinity.  Therefore calculation of the \emph{rate} in which $d$ approaches infinity gives a generalization of the original deficiency concept.

In the following theorem, a formula is derived for computing the generalized deficiency between two estimators from their MSE performance which explicitly computes the rate at which $d$ approaches infinity.
\begin{theorem}
\label{thm:power}
  Suppose the mean squared errors of two estimators $S_n$ and $T_n$ are given as
\[
\begin{split}
\text{MSE}(S_n)&=\frac{c}{n^r}+\frac{a}{n^{r+\delta}}+o\left(\frac{1}{n^{r+\delta}}\right)\\
\text{MSE}(T_n)&=\frac{c}{n^r}+\frac{b}{n^{r+\delta}}+o\left(\frac{1}{n^{r+\delta}}\right)
\end{split}
\]
Define $m=m(n)$ to be the sample size for which $\text{MSE}(T_m)$ equals (up to a second order term) $\text{MSE}(S_n)$.  Then the asymptotic deficiency of $T_n$ relative to $S_n$ is $d=m-n$ and satisfies
\[
\frac{d}{n^{1-\delta}}\longrightarrow\frac{b-a}{cr}
\]
\end{theorem}

In the next theorem, the deficiency of two estimators is calculated when the second-order term in the MSE expansion decreases at the rate $n^r\log n$ which is very close to the leading term of $n^r$.  Therefore the deficient index, $d$, will approach infinity at a faster rate indicating a larger discrepancy in the performance of the two estimators.

\begin{theorem}
\label{thm:log}
  Suppose the mean squared errors of two estimators $S_n$ and $T_n$ are given as
\[
\begin{split}
\text{MSE}(S_n)&=\frac{c}{n^r}+\frac{a}{n^r\log n}+o\left(\frac{1}{n^r\log n}\right)\\
\text{MSE}(T_n)&=\frac{c}{n^r}+\frac{b}{n^r\log n}+o\left(\frac{1}{n^r\log n}\right)
\end{split}
\]
Define $m=m(n)$ to be the sample size for which $\text{MSE}(T_m)$ equals (up to a second order term) $\text{MSE}(S_n)$.  Then the asymptotic expected deficiency of $T_n$ relative to $S_n$ is $d=m-n$ and satisfies
\[
d\left(\frac{\log n}{n}\right)\longrightarrow\frac{b-a}{cr}
\]
\end{theorem}

These formulas, combined with the results of Corollary 1 and equation \eqref{eq:var}, are used to derive the deficiency of infinite-order kernel estimators to the unsmoothed EDF under the assumptions $A(p)$, $B$, and $C$.  In the case of assumption $C$, the improvement in MSE performance is first-order, and therefore improvement in terms of efficiency, or ARE, is present.

\begin{corollary}
Let $\hat{F}_h(t)$ be as in Theorem \ref{thm:bias} and $\hat{F}$ be the empirical distribution function estimator.  Assume $F(t)\left(1-F(t)\right)\not=0$.
\begin{problist}
\item[(i)]  Suppose assumption $A(p)$ holds.  When $h\sim
an^{-\beta}$ where $a>0$ is constant and
$\beta=(2p+1)^{-1}$, the deficiency of $\hat{F}_h(t)$ relative to $\hat{F}(t)$ is
    \[
    \left(\frac{2af(t) \left(\int u\bar{K}(u)K(u)\,du\right)}{F(t)\left(1-F(t)\right)}\right)n^{\frac{2p}{2p+1}}
    \]
\item[(ii)]  Suppose assumption $B$ holds.  When $h\sim a/\log n$
where $a<2d$ is a constant, the deficiency of $\hat{F}_h(t)$ relative to $\hat{F}(t)$ is
    \[
    \left(\frac{2af(t) \left(\int u\bar{K}(u)K(u)\,du\right)}{F(t)\left(1-F(t)\right)}\right)\frac{n}{\log n}
    \]
\item[(iii)]  Suppose assumption $C$ holds.  When $h\le 1/b$ is constant, the deficiency of $\hat{F}_h(t)$ relative to $\hat{F}(t)$ is
    \[
    \left(\frac{2f(t) \left(\int u\bar{K}(u)K(u)\,du\right)}{F(t)\left(1-F(t)\right)}\right)n.
    \]
\end{problist}
\end{corollary}

\section{Bandwidth Selection}
\label{section:bandwidth}
We now present a simple bandwidth selection algorithm that requires very minimal computation and adapts to the specialized family of infinite-order kernels that is utilized in this paper.  The methods suggested in \cite{politis2003abc} for iid data and in \cite{Berg07} for censored data present an algorithm that automatically selects the optimal bandwidth in \emph{density} estimation.  Remarkably, these same algorithms can also be used to select the best bandwidth in CDF estimation.  Although the bias in estimating the CDF is smaller than the bias of the density estimators, the variance of the CDF estimator is also smaller than the variance of the density estimator.   This algorithm automatically adapts to the appropriate assumption $A(p)$, $B$, or $C$ and generates a bandwidth that is consistent for the ideal bandwidth given by Corollary \ref{cor:bias}.  The algorithm is also computationally light as well as being simple to describe, and we now proceed to describe it.

Let $\hat{\phi}$ be the natural estimate of the characteristic function given by
\[
\hat{\phi}(t)=\int_{-\infty}^\infty e^{itx}\,d\hat{F}(x)=\sum_{j=1}^ne^{itX_j}.
\]
In the context of censored data, $\hat{F}(x)$ in the above expression is replaced with the Kaplan-Meier estimator of the CDF.  The main key to the algorithm is finding when $\phi(t)\approx0$; more specifically, determining the smallest value $t^*$ such that $\phi(t)\approx 0$ for all $t\in(t^*,t^*+\varepsilon)$ for some pre-specified $\varepsilon$.  Then the estimate of the bandwidth is given by $\hat{h}=1/t^*$.  The formal algorithm is presented below.

\begin{quote}
  \textsc{Bandwidth Selection Algorithm}\\
  Let $C>0$ be a fixed constant, and $\varepsilon_n$ be a nondecreasing
  sequence of positive real numbers tending to infinity such that $\varepsilon_n=o(\log
  n)$.
  Let $t^*$ be the smallest number such that
  \begin{equation}
  \label{eq: alg}
|\hat{\phi}(t)|<C\sqrt{\frac{\log_{10} n}{n}}\qquad\text{ for all
}t\in (t^*,t^*+\varepsilon_n)
  \end{equation}
  Then let $\hat{h}=c/t^*$ where $c$ is the ``flat-top radius'' depicted in equation \eqref{eq:kappa}.
\end{quote}

The positive constant $C$ is irrelevant in the asymptotic theory, but is relevant for finite-sample calculations.  The main idea behind the algorithm is to determine the smallest $t$ such that $\phi(t)\approx 0$. In most cases this can be visually seen without explicitly computing the threshold in \eqref{eq: alg}.

\section{Boundary Correction and Standardization}
\label{section:boundary}
Vanilla versions of the kernel estimators for \emph{density} estimation break down when the support of the density is restricted to a subset of the real line.  For instance, in estimating the probability density function of data taken from an exponential distribution, most kernel estimators give substantial area to negative values even when it is known that the support of the density is nonnegative.  It is not too difficult to see that simple kernel estimators of the density will not be consistent at the boundary of the density's support; cf. \cite{silverman1986des}.  However, a simple remedy by reflection works well when the support is not too complex.  For instance when the support of the density is $[a,\infty)$, then the estimator
\begin{equation}
\label{eq:boundary}
\hat{\hat{f}}_h(x)=\left(\hat{f}_h(x)+\hat{f}_h(2a-x)\right)1_{[a,\infty)(x)}
\end{equation}
is consistent at the boundary point $a$ (\cite{silverman1986des}).

This problem, therefore, also carries over to the situation of estimating the CDF.  Indeed the EDF and Kaplan-Meier estimators do not suffer from this drawback, but the kernel smoothed versions do.  By integrating \eqref{eq:boundary}, we deduce a boundary-corrected version of the kernel-smoothed CDF estimator with the same formulation as \eqref{eq:boundary}.  For $t\in[a,\infty)$,
\[
\begin{split}
\hat{\hat{F}}_h(t)&=\int_{-\infty}^{t}\hat{\hat{f}}_h(x)\,dx\\
&=\int_{a}^{t}\left(\hat{f}_h(x)+\hat{f}_h(2a-x)\right)\,dx\\
&=\hat{F}_h(t)-\hat{F}_h(a)+\int_{2a-t}^{a}\hat{f}_h(x)\,dx\\
&=\hat{F}_h(t)-\hat{F}_h(a)+\hat{F}_h(a)-\hat{F}_h(2a-t)\\
&=\hat{F}_h(t)-\hat{F}_h(2a-t)
\end{split}
\]
and $\hat{F}_h(t)=0$ when $t<a$.  In the special case $a=0$, we have the simple formula
\[
\hat{\hat{F}}_h(t)=\left(\hat{F}_h(t)-\hat{F}_h(-t)\right)1_{[0,\infty)(t)}
\]

There is an additional issue that only affects higher-order kernel estimators and not second-order estimators.  Specifically, higher-order kernel estimators of the density are not necessarily nonnegative, which means higher-order kernels estimators of the CDF are not necessarily contained within the range $[0,1]$ or forced to be nondecreasing.  The natural remedy for these density estimators is to truncate negative estimates to zero and then renormalize the area to one.  When this is performed, the corresponding CDF estimator will be a valid CDF.  However this approach causes the kernel estimator of the CDF to lose its simplistic representation that is given in the right-hand side of \eqref{eq:Fhat}, so instead, a simple alternative standardization technique is suggested.  To insure the estimator is nondecreasing, $\hat{F}_h(t)$ is replaced by $\sup_{(-\infty,t)}\hat{F}_h(t)$, and to insure the range is between 0 and 1, $\max(\hat{F}_h(t),1)$ and $\min(\hat{F}_h(t),0)$ are invoked.

Replacing $\hat{F}_h(t)$ with $\sup_{(-\infty,t)}\hat{F}_h(t)$ is equivalent to replacing the estimator of the density $\hat{f}_h(x)$ with the truncated version $\hat{f}^+_h(x)=\max(\hat{f}_h(x),0)$ and then integrating the truncated density estimator from $-\infty$ to $t$.  Since $\hat{f}^+_h(x)$ has better MSE performance than the nontruncated counterpart $\hat{f}_h(x)$ \cite{politis95}, it follows that the nondecreasing estimator $\sup_{(-\infty,t)}\hat{F}_h(t)$ has better MSE performance than the original $\hat{F}_h(t)$.  Similarly, the MSE of the range restricted estimator produced from $\max(\hat{F}_h(t),1)$ and $\min(\hat{F}_h(t),0)$ will also be improved since it is known the CDF has a range bounded in [0,1].  This is formalized in the following corollary. 

\begin{corollary}
Let $\hat{F}_{h}(t)$ be as in Theorem 1.  A modified estimator is defined as
\[
\tilde{F}_{h}(t)=\max\left(\min \left(\sup_{(-\infty, t]}\hat{F}_{h}(t),0\right),1\right).
\]
Then it follows that
\[
\MSE\left(\tilde{F}_{h}(t)\right)\le \MSE\left(\hat{F}_{h}(t)\right)
\]
and $\tilde{F}_{h}(t)$ satisfies the necessary properties of a CDF.
\end{corollary}

\section{Simulations}
\label{section:simulations}
We evaluate the performance of the proposed infinite-order kernel estimators with the more traditional second-order kernel estimators and the EDF/Kaplan-Meier estimator.  Boundary correction, as described in Section \ref{section:boundary}, is applied to the estimators when appropriate.  As any choice of function $g(x)$ in \eqref{eq:kappa} will insure the ideal asymptotics of an infinite-order kernel, the selection of infinite-order kernels is quite large.  An easy choice for the function $g(x)$ is the straight line truncated at zero, i.e. $g(x)=(\frac{1-x}{1-c})^+$, which gives $\kappa$ a trapezoidal shape.   The simulations below considers this trapezoidal function $\kappa$ with $c=.75$.  

By making the flat-top function $\kappa(x)$ infinitely smooth, the resulting kernel via the Fourier transform will have tails that decay exponentially.  Therefore in situations in estimating the density with boundary conditions, the kernel derived from the infinitely smooth flat-top function is more close to having the desirable quality of being compactly supported than the kernel which is derived from the trapezoidal function.  One example of an infinitely smooth $\kappa(x)$ is \cite{mcmurry04}

\begin{equation}
\label{eq:flat}
\kappa(s)=
\begin{cases}
1 & \text{if } |s|<c\\
\exp\left( \frac{-b\exp\left(\frac{-b}{(|x|-c)^{2}}\right)}{(|x|-1)^{2}}\right) & \text{if }c<|x|<1\\
0 & \text{if }|x|\ge 1
\end{cases}
\end{equation}

which resembles and infinitely smooth trapezoid and is controlled by the two parameters $b$ and $c$.  In the simulations, we also used this function $\kappa$ for comparisons with the parameters $b=1$ and $c=.05$.  A plot of this $\kappa$ is given below.

\begin{figure}[h]
\centering
\includegraphics[height=5cm]{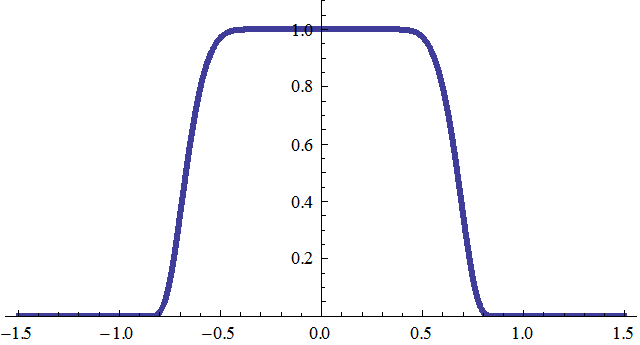}
\caption{Infinitely differentiable flat-top function \eqref{eq:flat}} with parameters $b=1$ and $c=.05$.
\end{figure}

This function is perfectly flat only from 0 to .05, but it is ``effectively'' flat from 0 to about .5.  Therefore the effective flat-top radius is taken to be .5, and it is this value that is used in the bandwidth selection algorithm described above in Section 4.

A slightly modified bandwidth selection algorithm was invoked that retains the function of the bandwidth algorithm described above.   The key in the bandwidth algorithm is to find the smallest value of $t^{*}$ so that $\hat{\phi}(t^{*})\approx 0$.  To automate this procedure, the value $t^{*}$ was chosen to be the first value for which $\hat{\phi}(t^{*})$ starts to level off.   

A Gaussian kernel is used in the second-order kernel estimator, and cross validation, as suggested in \cite{bowman1998bss}, is used to select the bandwidth for this estimator.  Estimates were simulated over 1000 realizations.

The first simulation study considers the estimation of a $N(0,1)$ CDF from iid data.  One may imagine the second-order Gaussian kernel estimator to do quite well in this context, but in fact the infinite-order kernel performs uniformly better.  MSE estimates are provided at three points ($t=-1.5,0,1.5$) and under two different sample sizes ($n=15,30$).

\ctable[pos=h, caption=Comparison of  the EDF with a Gaussian kernel estimator and two infinite-order
kernel estimators (trapezoid and smoothed trapezoid) on iid normal data]{rccccccccc} {\tnote[*]{MSE
values are blown up by $10^3$ for easier comparison.}} {
\FL & \mc{2}{$t=-1.5$} & & \mc{2}{$t=0$} & &\mc{2}{$t=1.5$} \NN
\cmidrule{2-3}\cmidrule{5-6}\cmidrule{8-9} 
$n$  & 15  & 30 &   & 15 & 30 &    & 15  & 30 \ML
{\color{black} $\text{MSE}_{\text{EDF}}$}\tmark[*] &   4.30 & 2.09 & & 16.29 & 8.73 & &  4.42 & 2.14 \NN
{\color{black} $\text{MSE}_{\text{Gauss}}$}\tmark[*] &   3.50 &  1.75 & &   13.02  &  7.20 & &  3.67 &   1.82 \NN
{\color{black} $\text{MSE}_{\text{trap}}$}\tmark[*] &{\bf  2.85} & {\bf 1.48} & & {\bf 11.72} & {\bf 6.49} & & {\bf 2.93} & {\bf 1.63} \NN
{\color{black} $\text{MSE}_{\text{smooth}}$}\tmark[*] & 2.95 & 1.55 & & 12.01 & 6.71 & & 3.06 & 1.69  \LL }

The second simulation study considers the estimation of a Weibull distribution with censored data.  Lifetime variables, the variables of interest, are simulated from a Weibull distribution with shape parameter 3 and scale parameter 1.5 and the censoring variables are independently drawn from a Weibull distribution with shape parameter 4 and scale parameter 3.   Since the support of the lifetime density is on the positive real line, the boundary correction of Section \ref{eq:boundary} is implemented.  MSE estimates are provided at three points ($t=.75,1.5,1.5$) and under two different sample sizes ($n=15,30$). Here again the infinite-order kernels consistently outperform the second-order kernel estimator and the Kaplain-Meier estimator in term of MSE performance.  In particular, the smoothed trapezoid is shown to perform well near the boundary point which can be attributed to its exponential tails making it more compactly supported.  

\begin{figure}[h]
\centering
\includegraphics{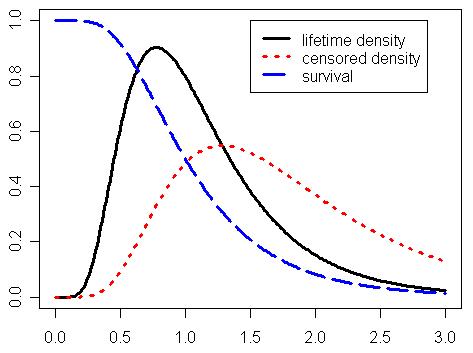}
\caption{Lifetime and censored Weibull densities considered in the simulations with a plot of the survival function also included.}
\end{figure}

\ctable[pos=H, caption=Comparison of  the EDF with a Gaussian kernel estimator and two infinite-order
kernel estimators (trapezoid and smoothed trapezoid) on censored Weibull data]{rccccccccc} {\tnote[*]{MSE
values are blown up by $10^3$ for easier comparison.}} {
\FL & \mc{2}{$t=.75$} & & \mc{2}{$t=1.25$} & &\mc{2}{$t=1.75$} \NN
\cmidrule{2-3}\cmidrule{5-6}\cmidrule{8-9} 
$n$  & 15  & 30 &   & 15 & 30 &    & 15  & 30 \ML
{\color{black} $\text{MSE}_{\text{EDF}}$}\tmark[*] &   6.47 & 3.51 & & 17.0 & 7.75 & &  12.0 & 5.62 \NN
{\color{black} $\text{MSE}_{\text{Gauss}}$}\tmark[*] &   5.45 & 2.84  & &   10.1  &  5.27 & &  8.56 &   4.11 \NN
{\color{black} $\text{MSE}_{\text{trap}}$}\tmark[*] &  5.83 & 2.70 & & {\bf 8.68} & {\bf 4.28} & & {\bf 9.32} & {\bf 4.06} \NN
{\color{black} $\text{MSE}_{\text{smooth}}$}\tmark[*] & {\bf 5.04} & {\bf 2.36} & & 9.81 & 4.85 & & 8.84 & 5.62  \LL }

\section{Discussion and Conclusions}
\label{section:discussion}

The proposed estimators have implications far beyond just providing a more accurate estimators of the CDF and survival function.  For instance, it is standard practice to compare the effects of two drugs based on their respected survival functions, but the cost of running clinical trials limits the sample size of the available data.  From the deficiency calculations of Section \ref{section:deficiency}, we see that the proposed estimators can produce the same results as the traditional Kaplan-Meier estimator yet with a significantly smaller sample size.

Another very standard use of the EDF is found in the bootstrap method.  In the smoothed bootstrap, data is drawn from a smoothed EDF, and when the estimator of the smoothed EDF is improved, the smoothed bootstrap is also improved to give more accurate inferences \cite{hall1989sab,polansky1997ksi}.  The bootstrap method is particularly beneficial when sample sizes are small, and therefore invoking infinite-order kernel estimators in this situation is often very natural.

Hazard function estimation on small samples can also be significantly be improved.  Hazard estimators, constructed from dividing a smoothed density estimate by a smoothed survival function, as in \cite{kim05}, have performance that is typically dictated by the convergence of the density estimator \cite{Berg07}.  However in small sample sizes, accurate estimation of the survival function is just as crucial as accurate estimation of the density.

The new infinite-order kernel estimators of the CDF and survival function is shown through analysis and demonstrated through simulations to be more accurate than the EDF and Kaplain-Meier estimators with significant improvements seen in small sample sizes and data from a distribution that has a rapidly decaying characteristic function.  Significant improvements in terms of an increase in efficiency is also produced by the new estimators when the characteristic function of the data is identically zero after some finite value.  Additionally, the bandwidth selection algorithm that accompanies the new estimator is computationally simpler with faster convergence rates than the cross-validation bandwidth selection algorithms used with finite-order kernels.

\appendix
\section{Technical Proofs}

\textsc{Proof of Theorem 1.}

From the following computation
\[
\E\left[\hat{F}_h(t)\right]=\frac{1}{n}\sum_{j=1}^n
\E\left[\bar{K}\left(\frac{t-X_i}{h}\right)\right],
\]
computing the bias of $\hat{F}_h(t)$ amounts to computing the bias
of $\bar{K}\left(\frac{t-X_i}{h}\right)$.  Starting with its
expectation, we have
\[
\begin{split}
\E\left[\bar{K}\left(\frac{t-X_i}{h}\right)\right]
&=\int_{-\infty}^\infty
\bar{K}\left(\frac{t-x}{h}\right)f(x)\,dx\\
&=\int_{-\infty}^\infty
\bar{K}\left(\frac{t-x}{h}\right)\,d F(x)\\
&=\underbrace{\bar{K}\left(\frac{t-x}{h}\right)F(x)\bigg|_{x=-\infty}^{x=\infty}}_{=0}+
\frac{1}{h}\int F(x)K\left(\frac{t-x}{h}\right)\,dx\\
&=\frac{1}{h}\int F(x)K\left(\frac{t-x}{h}\right)\,dx.
\end{split}
\]

If we define $K_h(t)=\frac{1}{h}K\left(\frac{t}{h}\right)$, then the
expectation above can be written in very simply as
\[
\E\left[\bar{K}\left(\frac{t-X_i}{h}\right)\right]=F\star K_h(t)
\]
where $\star$ denotes convolution.

To proceed, we will employ Fourier transform theory on
(mathematical) distributions, otherwise known as generalized
functions.  By invoking generalized functions, we can compute the
Fourier transform of not just the standard class of integrable
functions, but also many non-integrable functions like constants and
cumulative distribution functions.  This theory, in general, is very
technical and readers unfamiliar with the subject are referred to
\cite{beerends03} for a nice treatment of the subject.

As $K$ is the Fourier transform of $\kappa$, $\kappa$ is therefore
the inverse Fourier transform of $K$.  Through a simple change of
variables, we have
\[
\mathcal{F}^{-1}\left(K_h(t)\right)=\kappa(th)
\] where the notation $\mathcal{F}$ and $\mathcal{F}^{-1}$ will represent
the Fourier transform and its inverse.

Next we wish to derive the Fourier transform of the CDF $F(t)$. This
is the first generalized function that we encounter and its Fourier
transform involves the Dirac delta function, $\delta(s)$.  Using the
Heaviside step function $H(x)$ given by $H(x)=1(x>0)$, we rewrite
$F(t)$ as
\[
F(t)=\int_{-\infty}^t f(x)\,dx=\int_{-\infty}^\infty f(x)
H(t-x)\,dx=f\star H(t)
\]
Therefore the Fourier transform of $F(t)$ reduces to the product of
the Fourier transforms of $f(x)$ and $H(x)$; i.e.
\[
\begin{split}
\mathcal{F}\left(F(t)\right)
&=\phi(s)\left(\pi\delta(s)+\frac{1}{i s}\right)\\
&=\pi\phi(0)\delta(s)+\frac{\phi(s)}{i s}\\
&=\pi\delta(s)+\frac{\phi(s)}{i s}.
\end{split}
\]

We will now proceed with estimating the bias of $\hat{F}_h(t)$.
\[
\begin{split}
  \bias\left(\hat{F}_h(t)\right)&=K_h\star F(t)-F(t)\\
  &=\mathcal{F}\left(\mathcal{F}^{-1}\left(K_h\star
  F(t)-F(t)\right)\right)\\
  &=\mathcal{F}\left(\mathcal{F}^{-1}\left(K_h\right)\cdot\mathcal{F}^{-1}\left(
  F\right)-\mathcal{F}^{-1}\left(F\right)\right)\\
  &=\mathcal{F}\left(\left(\mathcal{F}^{-1}\left(K_h\right)-1\right)\mathcal{F}^{-1}\left(
  F\right)\right)\\
  &=\mathcal{F}\left(\left(\kappa(sh)-1\right)\left(\pi\delta(s)+\frac{\phi(s)}{i s} \right)\right)\\
  &=\mathcal{F}\left(\left(\kappa(sh)-1\right)\frac{\phi(s)}{i s} \right)-\pi\mathcal{F}\left(\left(\kappa(sh)-1\right)\delta(s)\right)\\
  &=\mathcal{F}\left(\left(\kappa(sh)-1\right)\frac{\phi(s)}{i s}
  \right)-\underbrace{\pi\mathcal{F}\left(\left(\kappa(sh)-1\right)\bigg|_{s=0}\right)}_{=0}\\
  &=\frac{1}{2\pi}\int_{|s|>1/h}\left(\kappa(sh)-1\right)\frac{\phi(s)}{i s}\, ds.
\end{split}
\]
The last equality comes from the flat-top property of $\kappa$
function; specifically, $\kappa(sh)=1$ for $|sh|\le 1$ implies
$\kappa(sh)-1=0$ for $|s|\le 1/h$.  Since $\kappa$ is bounded by
one, we have the following bound on the bias of $\hat{F}_h(t)$,
\[
\left|\bias\left(\hat{F}_h(t)\right)\right|\le
\frac{2}{2\pi}\int_{|s|>1/h}\frac{|\phi(s)|}{|s|}\,ds.
\]
We now bound the bias under the three assumptions $A(p)$, $B$, and
$C$.  Under assumption $A(p)$, we have
\begin{equation}
\label{eq:bias1}
\begin{split}
\int_{|s|>1/h}\frac{|\phi(s)|}{|s|}\,ds&=\int_{|s|>1/h}\frac{|s|^p|\phi(s)|}{|s|^{p+1}}\,ds\\
&\le h^{p+1}\int_{|s|>1/h}|s|^p|\phi(s)|\,ds\\
&=o(h^{p+1}).
\end{split}
\end{equation}
Under assumption $B$,
\begin{equation}
\label{eq:bias2}
\begin{split}
\int_{|s|>1/h}\frac{|\phi(s)|}{|s|}\,ds&
\le h\int_{|s|>1/h}|\phi(s)|\,ds\\
&\le h\int_{|s|>1/h}De^{-d|s|}\,ds\\
&\le \frac{D h}{e^{d/h}}\int_{|s|>1/h}e^{d(1/h-|s|)}\,ds\\
&= O\left(h e^{-d/h}\right).\\
\end{split}
\end{equation}
And under assumption $C$,
\begin{equation}
\label{eq:bias3} \int_{|s|>1/h}\frac{|\phi(s)|}{|s|}\,ds=0
\end{equation}
when $h\le 1$.  Therefore parts (i) through (iii) are proven from
equations \eqref{eq:bias1}, \eqref{eq:bias2}, and \eqref{eq:bias3}
respectively.

\textsc{Proof of Theorem 3.}

If the mean square errors are equal, up to a fraction of the sample size, then we have
\[
\frac{c}{n^r}+\frac{a}{n^{r+\delta}}+o\left(\frac{1}{n^{r+\delta}}\right)=
\frac{c}{m^r}+\frac{b}{m^{r+\delta}}+o\left(\frac{1}{m^{r+\delta}}\right)
\]
which implies
\[
\frac{1}{n^r}\left[c+\frac{a+o(1)}{n^\delta}\right]=
\frac{1}{m^r}\left[c+\frac{b+o(1)}{m^\delta}\right].
\]
Dividing through by $c$ and solving for $\frac{m}{n}$ gives
\[
\begin{split}
\frac{m}{n}=\left[1+\frac{b+o(1)}{cn^\delta}\right]^{1/r}\left[1+\frac{a+o(1)}{cm^\delta}\right]^{-1/r}.
\end{split}
\]
From the above expression, we see that $m/n\rightarrow 1$ and therefore $o(1/n)=o(1/m)$.  Using the approximation $(1+x)^s=1+sx+O(x^2)$ gives
\[
\begin{split}
\frac{m}{n}=1+\frac{b}{crn^\delta}-\frac{a}{crm^\delta}+o\left(\frac{1}{n^\delta}\right)
\end{split}
\]
Recalling $m=n+d$, we have
\[
\frac{d}{n}=\frac{b}{crn^\delta}-\frac{a}{crm^\delta}+o\left(\frac{1}{n^\delta}\right).
\]
Multiplying both sides of the above equation by $n^\delta$ gives
\[
\frac{d}{n^{1-\delta}}=\frac{b}{cr}-\frac{a}{cr}\left(\frac{n}{m}\right)^{\delta}+o\left(1\right)\longrightarrow \frac{b-a}{cr}.
\]

\textsc{Proof of Theorem 4.}

  The proof of Theorem \ref{thm:log} follows the same lines as the proof of Theorem \ref{thm:power} with $n^\delta$ replaced with $\log n$.

\bibliographystyle{plainnat}
\bibliography{deficiency3}

\end{document}